\documentclass[aps,prd,a4paper,twocolumn,amsmath,showpacs,superscriptaddress,nofootinbib,preprintnumbers]{revtex4-1}

\usepackage{graphicx,ulem}
\usepackage{longtable}
\usepackage{float}
\usepackage{dcolumn}
\usepackage{graphics,epsfig}
\usepackage{amsmath,amssymb,latexsym,mathrsfs}
\usepackage{bm}
\usepackage{color}
\usepackage{color}
\usepackage{subfigure}

\def\he4{$^4$He}
\def\h2{$^2$H}

\def\aap{\ref@jnl{A\&A}}                

\begin{document}
\preprint{NORDITA-2016-27,IFIC/16-22}

\title{On the improvement  of cosmological neutrino mass bounds}

\author{Elena Giusarma}
\email{egiusarm@andrew.cmu.edu}	
\affiliation{McWilliams Center for Cosmology, Department of Physics, Carnegie Mellon University, Pittsburgh, PA 15213, USA}

\author{Martina Gerbino}
\email{martina.gerbino@fysik.su.se}	
\affiliation{The Oskar Klein Centre for Cosmoparticle Physics, Department of Physics, Stockholm University, AlbaNova, SE-106 91 Stockholm, Sweden}
\affiliation{The Nordic Institute for Theoretical Physics (NORDITA), Roslagstullsbacken 23, SE-106 91 Stockholm, Sweden}

\author{Olga Mena}
\affiliation{IFIC, Universidad de Valencia-CSIC, 46071, Valencia, Spain}

\author{Sunny Vagnozzi}
\affiliation{The Oskar Klein Centre for Cosmoparticle Physics, Department of Physics, Stockholm University, AlbaNova, SE-106 91 Stockholm, Sweden}
\affiliation{The Nordic Institute for Theoretical Physics (NORDITA), Roslagstullsbacken 23, SE-106 91 Stockholm, Sweden}

\author{Shirley Ho}
\affiliation{McWilliams Center for Cosmology, Department of Physics, Carnegie Mellon University, Pittsburgh, PA 15213, USA}

\author{Katherine Freese}
\affiliation{The Oskar Klein Centre for Cosmoparticle Physics, Department of Physics, Stockholm University, AlbaNova, SE-106 91 Stockholm, Sweden}
\affiliation{The Nordic Institute for Theoretical Physics (NORDITA), Roslagstullsbacken 23, SE-106 91 Stockholm, Sweden}
\affiliation{Michigan Center for Theoretical Physics, Department of Physics, University of Michigan, Ann Arbor, MI 48109, USA}

\date{\today}

\begin{abstract}
The most recent measurements of the temperature and low-multipole polarization anisotropies of the Cosmic Microwave Background (CMB) from the Planck satellite, when combined with galaxy clustering data from the Baryon Oscillation Spectroscopic Survey (BOSS) in the form of the full shape of the power spectrum, and with Baryon Acoustic Oscillation measurements, provide a $95\%$ confidence level (CL) upper bound on the sum of the three active neutrinos $\sum m _\nu< 0.183$~eV, among the tightest neutrino mass bounds in the literature, to date, when the same datasets are taken into account. This very same data combination is able to set, at  $\sim70\%$~CL,  an upper limit on $\sum m _\nu$ of $0.0968$~eV, a value that approximately corresponds to the minimal mass expected in the inverted neutrino mass hierarchy scenario. If high-multipole polarization data from Planck is also considered, the $95\%$CL upper bound is tightened to $\sum m _\nu< 0.176$~eV. Further improvements are obtained by considering recent measurements of the Hubble parameter. These limits are obtained assuming a specific non-degenerate neutrino mass spectrum; they slightly worsen when considering other degenerate neutrino mass schemes. Current cosmological data, therefore, start to be mildly sensitive to the neutrino mass ordering. Low-redshift quantities, such as the Hubble constant or the reionization optical depth, play a very important role when setting the neutrino mass constraints. We also comment on the eventual shifts in the cosmological bounds on $\sum m_\nu$ when possible variations in the former two quantities are addressed.
\end{abstract}

\maketitle

\section{Introduction}\label{sec:intro}
Neutrinos are sub-eV elementary particles which, apart from gravity, only interact via weak interactions, decoupling from the thermal bath as ultra-relativistic states and constituting a \textit{hot} dark matter
component in our Universe. From neutrino mixing experiments we know
that neutrinos have masses, \textit{implying the first departure from
  the Standard Model (SM) of Particle Physics}\cite{Gonzalez-Garcia:2014bfa,Bergstrom:2015rba}. However, oscillation experiments are not sensitive to the absolute
neutrino mass scale; they only provide information on the squared mass
differences. In the minimal three neutrino scenario, the best-fit
value for the solar mass splitting is $\Delta m_{12}^2\simeq 7.5\times
10^{-5}$~eV$^2$ and for the atmospheric mass splitting is $|\Delta
m_{3i}^2|\simeq 2.45\times
10^{-3}$~eV$^2$~\cite{Gonzalez-Garcia:2014bfa}, with $i=1$ ($2$) for
the normal (inverted) mass scheme. Notice that the sign
of the largest mass splitting remains unknown, leading to two possible
hierarchical scenarios: \textit{normal} ($\Delta m_{31}^2>0$) and \textit{inverted} ($\Delta m_{32}^2<0$). In
the \textit{normal} hierarchy, $\sum m_\nu \gtrsim 0.06$~eV, while in the 
\textit{inverted} hierarchy, $\sum m_\nu \gtrsim 0.10 $~eV, with $\sum
m_\nu$ representing the total neutrino mass.

Neutrinos, as \textit{hot} dark matter particles, possess large
\textit{thermal} velocities, clustering only at $k<k_{fs}$, i.e. at scales below the
\textit{neutrino free streaming wavenumber $k_{fs}$}, and suppressing structure
formation at $k>k_{fs}$~\cite{Bond:1980ha,Hu:1997mj}. The presence of massive neutrinos
also affects the CMB, as these particles 
may become non-relativistic around the photon decoupling period. In particular they change the matter-radiation
equality causing a small shift in the peaks of the CMB and a mild increase of their heights due to the
Sachs-Wolfe effect. In addition, current CMB experiments allow one to explore the impact of massive neutrinos at small scales (\textit{i.e.} at high multipoles $\ell$), because they
are sensitive to the smearing of the acoustic peaks caused by the gravitational lensing of CMB photons \cite{Lewis:2006fu}. Cosmology can
therefore \textit{weigh} relic neutrinos. 
Recent studies dealing with the cosmological constraints on $\sum m_\nu$ have reported $95\%$~CL upper bounds
of $0.754$~eV and $0.497$~eV from Planck temperature anisotropies and
Planck temperature and polarization measurements, respectively~\cite{DiValentino:2016ikp}. In order
to improve these CMB neutrino mass limits, additional information from additional dark
matter tracers and/or other geometrical standard rulers are
needed. Current cosmological upper bounds on $\sum m_\nu$, 
which combine CMB temperature and polarization
anisotropies measurements with different observations of the large scale structure of the Universe, 
range from $0.12$~eV to $0.13$~eV at
$95\%$~CL~\cite{Palanque-Delabrouille:2015pga,DiValentino:2015sam,DiValentino:2015wba,Cuesta:2015iho}.
These limits are extremely close to the predictions from
neutrino oscillation experiments in the \textit{inverted} hierarchical spectrum. However, we note that the strongest limits among the ones quoted above have been obtained by employing Planck polarization measurements at small scales \cite{Ade:2015xua}, which could be affected by a small residual level of systematics ~\footnote{We also note that, even though results from \cite{Palanque-Delabrouille:2015pga} are shown in combination with Planck temperature (\textit{i.e.} without small scale polarization), they come from a frequentist analysis. As detailed below, we are going to show results obtained within the bayesian framework. As a result, a direct comparison between our limits and  \cite{Palanque-Delabrouille:2015pga} is hard to assess.}.

Here we follow a more conservative approach. We exploit the effect of the neutrino masses in galaxy
 clustering, focusing on the full 3D galaxy power spectrum shape 
from the Baryon Oscillation Spectroscopic Survey
(BOSS)~\cite{Dawson:2012va} Data Release 9 (DR9)~\cite{Ahn:2012fh}
 (which is among the largest sets of galaxy spectra publicly
 available to date), in combination with the Planck CMB 2015 full data in temperature, complemented with large scale polarization measurements \cite{Adam:2015rua}. This is our baseline combination. When we combine two data sets - independent large scale structure measurements in
 the form of Baryon Acoustic Oscillations (BAO) and different priors on the
 Hubble parameter~\cite{Riess:2016jrr, Efstathiou:2013via} -  the minimal value of the mass expected in the inverted
 neutrino mass hierarchical scenario (see text below for a definition of minimal value in this context), $\sum m _\nu=0.0968$~eV, is
 excluded up to a significance of $90\%$~CL. This indicates that current cosmological measurements show a
 mild preference for the region of the parameter space corresponding
 to the normal hierarchical scheme for the neutrino mass
 eigenstates. Moreover, cosmological data start to show differences in the
 neutrino mass bounds for the different possible neutrino mass
 schemes. We also illustrate the very important role played by
 low redshift observables and how they affect the limit quoted above. 

\section{Analysis and data}\label{sec:method}
In the following section, the cosmological model we assume is the standard $\Lambda$CDM
scenario, described by its usual six parameters, plus the sum of the
neutrino masses $\sum m_\nu$. In particular, the model
parameters are the baryon $\Omega_b  h^2$ and the cold dark matter
$\Omega_c h^2$ physical mass-energy densities, the ratio between the sound
horizon and the angular diameter distance at decoupling $\Theta_{s}$,
the reionization optical depth $\tau$,  the scalar spectral index
$n_s$ and the amplitude of the primordial spectrum $A_{s}$. We follow
here the Planck $\Lambda$CDM model assumption of two massless neutrino
states and a massive one. In addition, we also present the
limits obtained when assuming one massless plus two massive neutrino
states instead. We compare these bounds to those in the three
degenerate massive neutrino scheme. In doing so, we are motivated by
the fact that current limits on $\sum m_\nu$ start excluding the
degenerate region at a high significance. As a result, it is timely to
investigate the impact of our assumptions on how the total mass is
distributed among the massive eigenstates. 
More detailed analyses will be carried out in an upcoming work \cite{inprep}.

Measurements of the CMB anisotropy temperature, polarization,
and cross-correlation spectra are exploited with the full
Planck 2015 data release~\cite{Adam:2015rua,Ade:2015xua}. We present
results arising from the combination of the full temperature data with
the large scale polarization measurements (i.e.  the Planck low-$\ell$ multipole likelihood that extends from
$\ell$ = 2 to $\ell$ = 29), referring to
it as \textit{Planck TT}. When combined with DR9, we refer to it as our \textit{Base} dataset. Furthermore, we also consider for the sake of comparison the addition of the small-scale 
polarization and cross-correlation spectra as measured by the Planck
High Frequency Instrument (HFI), which in the following will be named \textit{Planck
  pol}. We refer to the combination of \textit{Planck pol} and DR9 as \textit{Basepol}. We analyze Planck CMB datasets, making use of the Planck likelihood
\cite{Aghanim:2015xee}. With respect to the different parameters involved
in the CMB foreground analyses, we vary them following
Refs.~\cite{Ade:2015xua,Aghanim:2015xee}. Because of a
  possible residual level of systematics in the coadded polarization
  spectra at high-multipoles, the Planck Collaboration suggests treating the full temperature and polarization results with caution~\cite{Ade:2015xua}. For
  this reason, we shall assume the \textit{Planck TT} as our CMB
  baseline data  and provide results from \textit{Planck pol} for the sake of comparison with other recent works~\cite{Palanque-Delabrouille:2015pga,Cuesta:2015iho}.

Together with Planck CMB temperature and polarization measurements, we
exploit here the DR9 CMASS sample of galaxies~\cite{Ahn:2012fh}, 
as previously done in Refs.~\cite{Zhao:2012xw,Giusarma:2013pmn}. 
This galaxy sample contains $264\,283$ massive galaxies over 3275\,deg$^2$ of the sky. The
redshift range of this galaxy sample is $0.43<z<0.7$, with a mean
redshift of $z_{\rm eff}=0.57$. The measured galaxy power spectrum $P_{\rm meas}(k)$ is identical to the one exploited for the BAO analyses~\cite{Anderson:2012sa}, and it is
affected by several systematic uncertainties, as carefully studied
in~\cite{Ross:2012qm,Ross:2012sx}. Following this previous work, we add an extra free
parameter to account for systematics in the measured power spectrum:
$P_{\rm meas}(k) = P_{\rm meas,w}(k)-S[P_{\rm meas,nw}(k)-P_{\rm
meas,w}(k)]$, where $P_{\rm meas,w}(k)$ is the measured power
spectrum after accounting for systematic uncertainties, $P_{\rm
  meas,nw}(k)$ refers to the measured power spectrum without these
effects, and $S$ is an extra nuisance parameter that will  be
marginalized over. Previous works~\cite{Zhao:2012xw,Ross:2012sx}
have applied a gaussian prior with a standard
deviation of $0.1$ to the $S$ parameter , based on the mocks of Ref.~\cite{Ross:2012qm}. 
Here we follow the same assumption for the systematics parameter
$S$.

\begin{figure*}[!htb]
\vspace{-0.1cm}
\centering
\includegraphics[width=1.5\columnwidth]{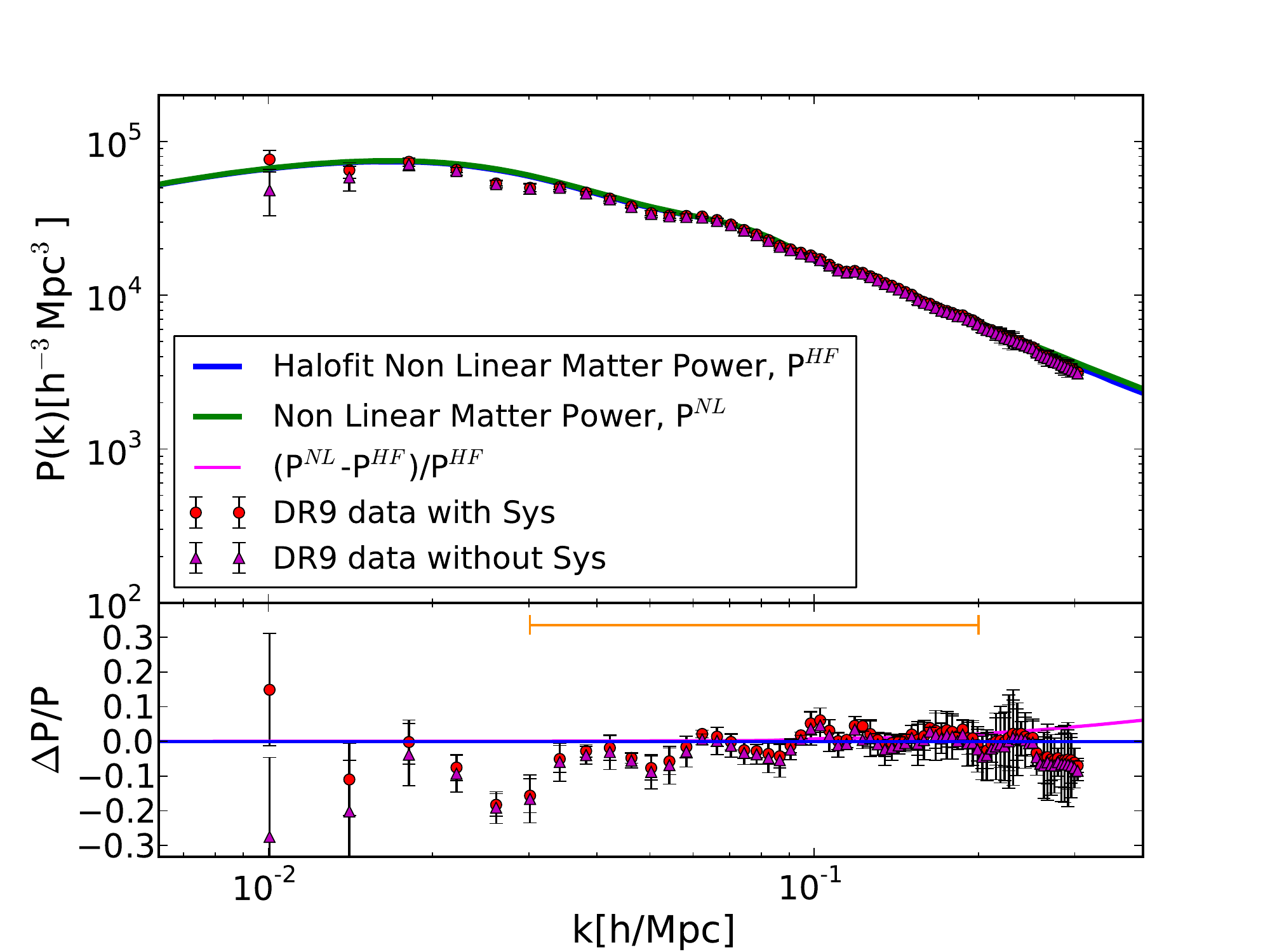}  
\caption{\textit{Top}: 
Non-linear matter power spectrum computed using the HaloFit method with the CAMB code~\cite{Lewis:1999bs} 
(blue line) and the Coyote emulator (green line) of Kwan et al.~(2015)~\cite{Kwan:2013jva}
at z=0.57 for the $\Lambda$CDM best-fit parameters from \textit{Planck TT} 2015
data.
Data points are the clustering measurements from the BOSS Data Release
9 (DR9) CMASS sample. The error bars are computed from the diagonal
elements $C_{ii}$ of the covariance matrix. We also illustrate the data after applying a maximal correction for systematics, i.e. $S=1$; see text for details.
\textit{Bottom}: Residuals with respect to the non linear model with HaloFit. The orange horizontal line indicates the $k$ range used in our analysis.}
\label{fig:power}
\end{figure*}

The expectation value of the matter power
spectrum requires a previous convolution of the true power spectrum with the window
functions. These functions describe the correlation of the data at different scales
$k$ due to the survey geometry, to be convolved with the theoretical power
spectrum, \textit{i.e.} the predicted power spectrum as a function of cosmological parameters extracted at each step of the Monte Carlo. The model galaxy power spectrum $P^g_{\textrm{th}}(k)$ is computed as  
$P^g_{\textrm{th}}(k, z) = b_{\rm HF}^{2 }P^{\rm m }_{\rm HF\nu}(k; z)
+ P^{\rm s}_{\rm HF}~$, where $P^\mathrm{m}_{\rm HF\nu}$ is the model matter power spectrum, with the scale independent parameters $b_{\rm HF}$ and $P^{\rm s}_{\rm HF}$ referring to
the bias and the shot-noise contribution respectively; see
\cite{Zhao:2012xw} for their adopted priors. The subscript HF refers
to the HaloFit prescription. Indeeed, we obtained the theoretical matter
power spectrum by making use of the HaloFit method~\cite{Smith:2002dz,Takahashi:2012em}, following corrections for
modeling in the presence of massive neutrinos from \cite{Bird:2011rb}. In order to reduce the impact of non
linearities, we adopt the conservative choice of a maximum wavenumber 
of $k_{\rm{max}}=0.2$~$h$/Mpc. As we can see in Fig.~\ref{fig:power}, this region is safe against very
large and uncertain non linear corrections in the modeled theoretical
spectra. Furthermore, this choice is also convenient for comparison
purposes with recent related work, see e.g.~\cite{Cuesta:2015iho}.

Even if the well-known degeneracy between the neutrino mass and the Hubble
constant $H_0$~\cite{Giusarma:2012ph} can be improved with large scale structure
data, an additional prior on the Hubble parameter helps in further
pinning down the cosmological neutrino mass limits, as we shall see  in
the following sections. A recent study~\cite{DiValentino:2015sam} has shown that the
choice of the low redshift priors plays a crucial role when constraining $\sum
m_\nu$. In particular, when considering the \textit{Planck pol} and Hubble
constant measurements, the $95\%$~CL limit on $\sum m_\nu$ was
between $0.34$ and $0.18$~eV, depending on the value of $H_0$
used in the analyses. 
Therefore, we also consider the combination of Planck and DR9 measurements with three different $H_0$
priors, two of which arise from the re-analysis carried out in
Ref.~\cite{Efstathiou:2013via}, consisting of a lower estimate 
($H_0=70.6\pm 3.3$ km s$^{-1}$ Mpc$^{-1}$) and a higher estimate
($H_0=72.5\pm 2.5$ km s$^{-1}$ Mpc$^{-1}$) of the Hubble
parameter. 
The third $H_0$ measurement used here relies on the recent measurement
reported in Ref.~\cite{Riess:2016jrr}, $H_0=73.02\pm 1.79$ km s$^{-1}$
Mpc$^{-1}$, by means of observations of Cepheids variables from the
Hubble Space Telescope (HST) in a number of novel host galaxies. 
This new estimate of the Hubble constant from HST has reduced its previous
uncertainty (see e.g ~\cite{Riess:2011yx}) to the $2.4\%$ level, and as
has lead to the tightest neutrino mass constraints, which we
presente in the following sections. Notice that these results should
be considered as the less conservative ones obtained in our study,
since the value of $H_0 = 73.02 \pm 1.79$~km s$^{-1}$
Mpc$^{-1}$ is $3\sigma$ higher than the Planck CMB $H_0$
value. Unaccounted systematic effects for both measurements may be
the origin for this discrepancy. Nevertheless,
the findings of \cite{Efstathiou:2013via}, yielding $H_0=72.5\pm 2.5$
km s$^{-1}$ Mpc$^{-1}$, are also in tension with the Planck Hubble
constant estimates (albeit in this case the tension is milder, at the
$2\sigma$ level). For the lower estimate of $H_0=70.6\pm 3.3$ km s$^{-1}$ Mpc$^{-1}$ from
\cite{Efstathiou:2013via}, the tension is much less
significant. In order to illustrate the very
important role of the Hubble constant prior, and how its choice may bias significantly
the results concerning the neutrino mass ordering preferred by current
cosmological data, we will present the neutrino mass limits for the
three possible cases described above and named \textit{H070p6}, \textit{H072p5} and
\textit{H073p02}. 

To provide a comparison with previous limits in the literature, we combine the Planck CMB plus
DR9 large scale structure measurements with additional large scale
structure information in the form of the BAO clustering signature. We
exploit BAO data results at $z_{\rm eff}=0.106$ from the 6dF Galaxy Survey
(6dFGS)~\cite{Beutler:2011hx},  $z_{\rm eff}=0.44$, $0.6$ and $0.73$ from the
WiggleZ Dark Energy Survey~\cite{Blake:2011en} and $z_{\rm eff}=0.32$ from BOSS Data Release 11 LOWZ 
sample~\cite{Anderson:2013zyy}.
The combination of these three datasets will be referred to as
\textit{BAO}~\footnote{The authors of Ref.~\cite{Cuesta:2015iho} exploit
both the LOWZ and the CMASS BOSS
measurements and, therefore, the impact of the BAO data is larger for that case \cite{2010JCAP...07..022H}.}.

In order to derive the cosmological constraints on the parameters, we
use the Monte Carlo Markov Chain (MCMC) package
\texttt{cosmomc} \cite{Lewis:2002ah,Lewis:2013hha}, using the Gelman
and Rubin statistics~\cite{An98stephenbrooks} for the convergence of the generated chains. 

\section{Results on $\sum m_\nu$}\label{sec:mnu}
\begin{figure}[!htb]
\vspace{-0.1cm}
\centering
\includegraphics[width=8.5cm]{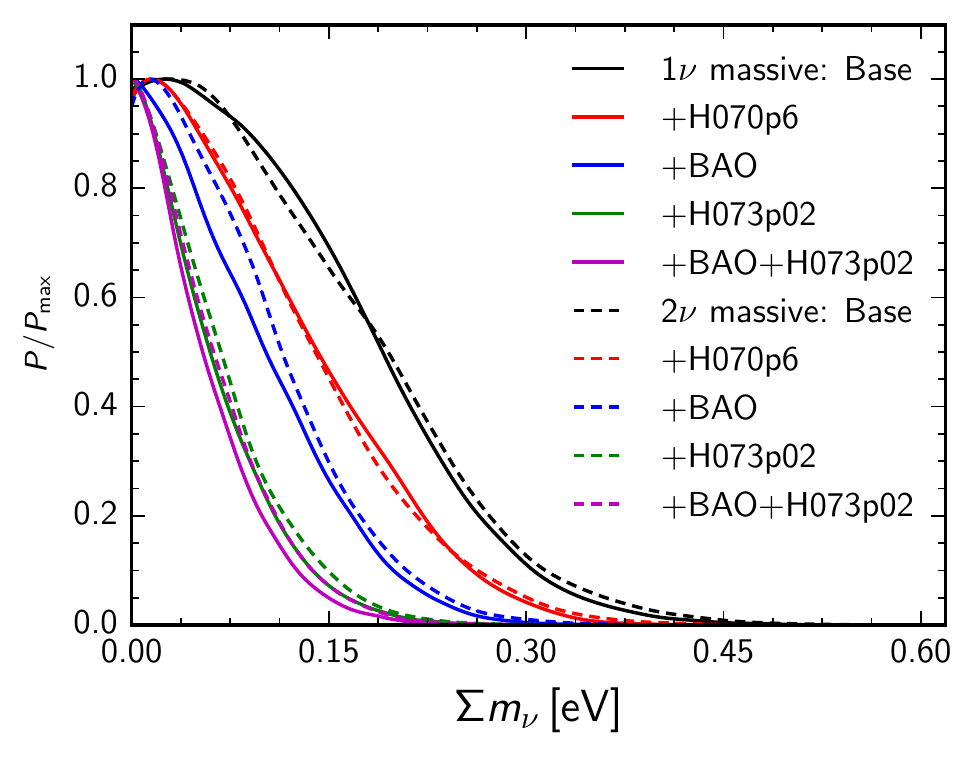}  
\caption{One-dimensional posterior probability for $\sum m_\nu$
  for the \textit{Base} combination, which consists of
  \textit{Planck TT} and DR9  galaxy clustering measurements, and also
  combined with other possible data sets. Both the one (solid) and the two (dashed) massive neutrino cases are illustrated.}
\label{fig:probfigmnu}
\end{figure}

In the following, we present the limits on the
total neutrino mass $\sum m_\nu$, imposing $\sum
m_\nu >0$, as in
Refs.~\cite{Palanque-Delabrouille:2015pga,Cuesta:2015iho}, and
mainly focusing on the case with one massive and two massless neutrino
states, although we also quote and discuss the bounds for other
possible assumptions concerning the neutrino mass
spectrum. Even if the cases of one massive plus two
massless and two massive plus one massless can be regarded as an
approximation of the normal (with $m_3\gg m_1\simeq m_2$) and inverted
(with $m_1\simeq m_2 \gg m_3$) hierachy, respectively, an assessment
about the preference of one scheme with respect to others is beyond
the scope of the present work. Our goal is to highlight possible
variations of the limits on $\sum m_\nu$ when assuming different mass
schemes as a proxy of the sensitivity of cosmological probes to the
neutrino mass hierarchy.
Table~\ref{tab:tabmnutt} shows our results in terms of the $95\%$~CL
upper limits on $\sum m_\nu$ (in eV) and the mean values, together with
their associated $95\%$~CL 
errors of the low redshift observables
$\tau$ and $H_0$, for the \textit{Base} combination of \textit{Planck
  TT} plus
DR9 galaxy clustering measurements, together with other external data
sets. The three possible neutrino mass spectral cases are
illustrated. Notice that the tightest limits are obtained
for the case of one massive state, for which we obtain a slight improvement of $\Delta
\chi^2\simeq 2$ with respect to the other two mass scenarios when
considering the \textit{H073p02} prior in the analyses. This is due to the fact that a pure
radiation component in the universe at late times alleviates the
tension between local
and high-redshift estimates of the Hubble constant. The associated
one-dimensional posterior probabilities for $\sum m_\nu$ (in eV) are depicted in
Fig.~\ref{fig:probfigmnu}, where we show the comparison among
different data sets for both the \textit{one} and the \textit{two}
massive neutrino assumptions.

The tightest $95\%$~CL upper bound on the neutrino mass is obtained for the \textit{Base}
combination together with the BAO and the \textit{H073p02} data sets, which notably
help in improving the neutrino mass limits, as we find  $\sum
m_\nu<0.125$~eV, $\sum m_\nu<0.135$~eV and $\sum
m_\nu<0.139$~eV for the one massive, two massive and degenerate
spectrum cases, respectively. According to the latest results on neutrino oscillation physics from
global fits~\cite{Gonzalez-Garcia:2014bfa}, in the inverted hierarchy,
the minimal value allowed for the total neutrino mass is $\sum m_\nu=0.0968$~eV. We 
choose to define the minimal value as the one obtained by setting the 
lightest eigenstate to zero and considering the $3\sigma$ 
allowed ranges of the mass differences from~\cite{Gonzalez-Garcia:2014bfa} (see~\cite{Gerbino:2015ixa} for a more detailed discussion about the definition of the minimal mass value).
The minimal neutrino mass in the inverted hierarchy scenario ($0.0968$~eV) is excluded by
the \textit{base} combination, BAO and the \textit{H073p02} prior on the Hubble constant
 at $88\%$~CL. This exclusion becomes less significant when
 the one massive neutrino scenario assumption is relaxed and the hot dark matter energy
 density is shared by either two or three massive neutrino states,
 cases for which we can exclude the region above $\sum m_\nu=0.0968$~eV at
 more modest significance levels ($85\%$ and $84\%$, respectively). 

\begin{figure}[tb!]
\vspace{-0.1cm}
\centering
\includegraphics[width=8.5cm]{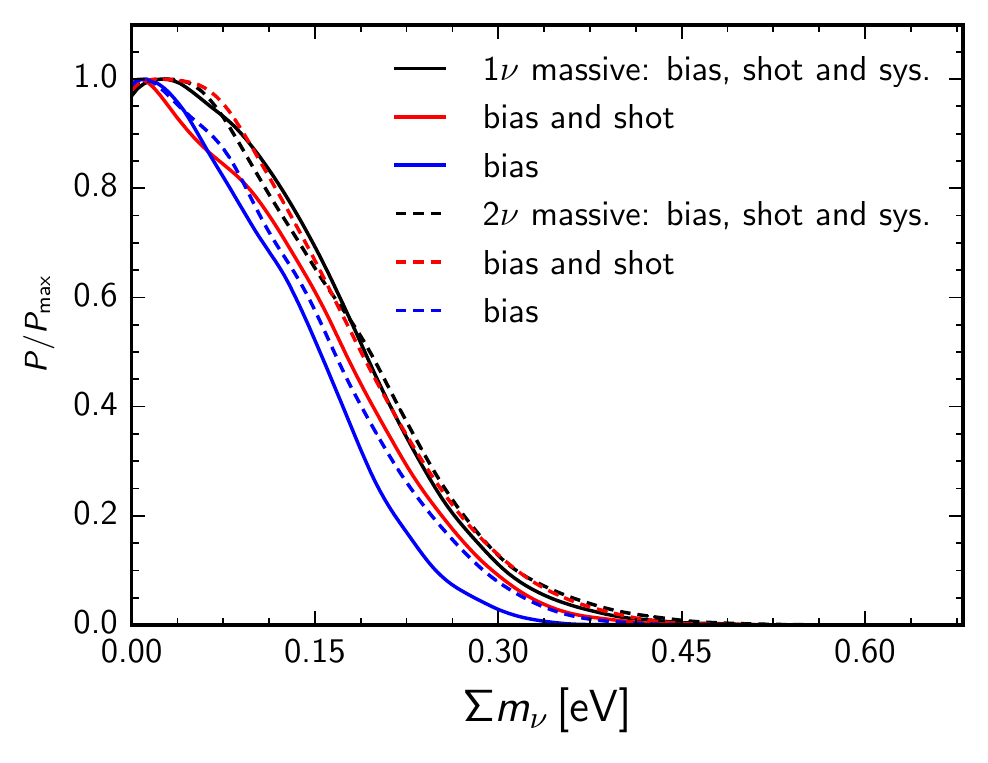}  
\caption{As in Fig.~\ref{fig:probfigmnu} but focusing on the
  \textit{Base} combination only. Different curves show the impact of marginalizing over bias, shot noise and systematics; see text for details.}
\label{fig:probfigmnu2}
\end{figure}

Notice that the bounds noted above are among the strongest ones in the
 literature, and are derived using \textit{Planck TT} data only.
The tightest limit quoted in
   Ref.~\cite{Palanque-Delabrouille:2015pga}, obtained with a
   different large scale structure tracer, namely, the Lyman
   $\alpha$ forest power spectrum, $\sum m_\nu <0.12$~eV at $95\%$~CL, is very
   close to our bound, as well as the bound $\sum m_\nu <0.13$~eV at 95\% from \cite{Cuesta:2015iho}. However, we recall here that our limit $\sum
m_\nu<0.125$~eV is obtained with a bayesian analysis and with \textit{Planck TT} data
 only. 
Furthermore, our limits arise from a conservative
 analysis accounting for all the possible factors which, in principle, may
 drastically reduce the constraining power of the DR9 large scale structure data. This
 can be noticed in the results depicted in
 Fig.~\ref{fig:probfigmnu2}, which shows the one dimensional probability distribution for $\sum
m_\nu$ considering the \textit{base} dataset for both the
\textit{one} and the \textit{two} massive schemes resulting from
different marginalizations (bias only, bias and shot-noise only, and,
finally, with systematics also included). Notice that, while
 systematic corrections do not affect our results, shot-noise
 contributions have a major impact. Indeed, we get $\sum
m_\nu<0.220\,\mathrm{eV}$ at 95\% CL for the \textit{Base} combination
 of datasets without the shot-noise contribution, whereas we get $\sum
m_\nu<0.269\,\mathrm{eV}$ at 95\% CL when marginalizing over bias and shot-noise for the same data.
On the other hand, the limits quoted above rely on the \textit{one} massive neutrino assumption
as well as on the addition of the recently derived \textit{H073p02}
prior. We shall comment on the impact of these two factors below.

For the sake of comparison with previous results in the literature~\cite{Palanque-Delabrouille:2015pga,Cuesta:2015iho},
we also present here the constraints obtained when high-multipole polarization data are also included in the analyses. Table~\ref{tab:tabmnupol} shows the $95\%$~CL upper bound on $\sum
m_\nu$ (in eV) and the mean values, together with
their associated $95\%$~CL 
errors, of the low redshift observables
$\tau$ and $H_0$, arising from the
analyses of \textit{Planck pol} plus DR9 data (combination named as
\textit{basepol}). 
In general, the results follow the same pattern than the ones obtained
before in the absence of polarization measurements: the combination of \textit{basepol} plus the
\textit{H073p02} prior sets an upper $95\%$~CL bound on $\sum m_\nu$ of
$0.125$~eV in the one massive neutrino case. If BAO measurements are
added to the former combination, the $95\%$~CL upper bounds
on the total neutrino mass reach very tight limits, corresponding to $\sum m_\nu <
0.123$~eV,  $\sum m_\nu < 0.113$~eV and $\sum m_\nu <
0.124$~eV in the \textit{one} massive, \textit{two}
massive and \textit{degenerate} neutrino mass spectra,
respectively. The minimal neutrino mass in the inverted hierarchy scenario ($0.0968$~eV) is excluded by
the \textit{basepol} combination, BAO and the \textit{H073p02} prior on the Hubble constant
 at $90\%$~CL in the \textit{one} massive neutrino scenario. In the
two massive and degenerate neutrino scenarios the significance of the exclusion is $91.8\%$
 and $88.6\%$~CL, respectively. 

As previosuly stated, there is a tension between  the
\textit{H072p5} and \textit{H073p02} priors and the Planck estimates
of the Hubble constant. Given the well-known degeneracy between
$H_0$ and $\sum m_\nu$~\cite{Giusarma:2012ph}, this tension should be carefully examined when
interpreting the $\sum m_\nu$ limits obtained here. Notice that the
highest $H_0$ priors (\textit{H072p5}
and \textit{H073p02}) lead to the tightest neutrino
mass constraints here; therefore, these limits should be regarded
as our less conservative bounds. As a rough test, we can compare the $\Delta \chi^2$ with respect to the \textit{Base} model: we get $\Delta \chi^2=4$ and 8 when \textit{H072p5}
and \textit{H073p02} priors are employed, respectively. Future accurate local determinations of
the Hubble constant could be shifted to larger (smaller) values,
 tightening (softening) the constraints found here. Another low-redshift observable which has a large
   impact on the cosmological bounds on $\sum m_\nu$ is the
   reionization optical depth, $\tau$. A recent estimation of the
   optical depth from the Planck collaboration, based on refined
   analyses of the polarization data of the Planck HFI on large
   angular scales, gives $\tau=0.055\pm 0.009$~\cite{Aghanim:2016yuo}, value which is in a better agreement with astrophysical measurements of Lyman-$\alpha$ emiters or high-redshift  
   quasars~\cite{Choudhury:2014uba,Mesinger:2014mqa,Mitra:2015yqa}
   than previous CMB estimates. This new value of $\tau$ will strengthen the bounds quoted
   here, see e.g.~\cite{DiValentino:2015sam}, as a smaller value of $\tau$ is translated into a smaller
   clustering amplitude. To avoid further reductions of the
   clustering amplitude, the contribution from massive neutrinos must be reduced.

A highly motivating effect is the fact that, even if there exists a small difference in the bounds for the
three possible neutrino mass schemes, they are indeed different, implying that present cosmological measurements are mildly sensitive to the
distribution of hot dark matter and radiation at late times. This effect can be understood by means of the suppression induced by
  relativistic and non-relativistic neutrino species in the growth of
  matter fluctuations. While in the two
  massive (or in the degenerate) neutrino scenario, there is only one (none) neutrino
  species which is relativistic
  today; in the one massive neutrino scheme, two neutrino
  species are relativistic at the current epoch~\footnote{As
  previously stated, this could be regarded as an approximation of normal and inverted hierarchical distribution of mass among the massive eigenstates.}. In the two-massive case, the
  power spectrum of matter fluctuations is suppressed due to the
  existence of two hot dark matter particles and one relativistic
  state that does not contribute to clustering.  In the one-massive case,
  the suppression of the growth of matter perturbations is larger, as
  there are two massless states that will not contribute to
  clustering. In addition, the free streaming wavenumber $k_{fs}$ associated with the
  massive state is larger than in the two massive or degenerate
  scenarios; therefore, there are more available modes to be exploited with the neutrino
  signature imprinted, benefiting as well from smaller
  error bars. Notice that, for the very same reasons, the different
  distribution of the total mass $\sum m_\nu$ among the massive eigenstates also affects the shape of the CMB power spectra, mainly due to the gravitational lensing effects.

This should be regarded as an example of how close we are to the limit
at which the usual approximations followed when exploring the
$\Lambda$CDM+ $\sum m_\nu$ scenario with cosmological probes become relevant. While statistical fluctuations could originate
some tiny shifts in the neutrino mass limits obtained in the three
possible neutrino mass spectrum scenarios~\footnote{By requiring a
convergence level (quantified by the Gelman and Rubin statistics $R$~\cite{An98stephenbrooks}) of $R-1\sim0.01$, the contribution from
statistical fluctuations can be roughly estimated to be a few percent the limits quoted in Tables~\ref{tab:tabmnutt} and
\ref{tab:tabmnupol}.}, current cosmological data already exclude the
degenerate region (with $\sum m_\nu$ well above 0.2 eV) at a high
significance, cornering the validity of the standard degenerate
neutrino assumption. Analyses involving an accurate inclusion of information
from oscillation measurements, along with a statistical model
comparison able to assess the preference for a hierarchy, become pressing, and will be performed
elsewhere~\cite{inprep} (see some previous related work in
Ref.~\cite{DeBernardis:2009di}). In addition, the forecasted sensitivity to $\sum m_\nu$ from future surveys makes the more rigorous approach outlined above unavoidable.

\begin{table*}[h!]

\begin{tabular}{|c||c|c|c|c|c|c|c|c|c|}
\hline
Dataset	&\multicolumn{3}{c|}{\textit{1 massive state}} &\multicolumn{3}{c|}{\textit{2 massive states}} &\multicolumn{3}{c|}{\textit{Degenerate spectrum}}\\
\hline
	&$\sum m_\nu$	&$\tau$	&$H_0$
 &$\sum m_\nu$	&$\tau$	&$H_0$ &$\sum m_\nu$	&$\tau$	&$H_0$\\

\hline\hline
\textit{Planck TT}& $<0.662$	&$0.080^{+0.038}_{-0.037}$
&$65.5^{+3.7}_{-4.3}$	&$< 0.724$	&$0.081^{+0.039}_{-0.038}$
&$65.4^{+4.2}_{-5.3}$ &$< 0.720$	&$0.080^{+0.038}_{-0.037}$	&$65.6^{+4.2}_{-5.7}$	\\	
base	&$< 0.269$	&$0.073\pm0.037$	&$66.8^{+2.1}_{-2.3}$	&$< 0.281$	&$0.073^{+0.037}_{-0.036}$	&$66.8^{+2.1}_{-2.3}$
&$< 0.297$	&$0.073^{+0.036}_{-0.037}$	&$66.8^{+2.1}_{-2.3}$	\\
base+BAO	&$< 0.183$	&$0.075\pm0.036$	&$67.5^{+1.4}_{-1.6}$	&$< 0.191$	&$0.075^{+0.037}_{-0.036}$	&$67.6^{+1.4}_{-1.6}$
&$< 0.202$	&$0.075^{+0.037}_{-0.038}$	&$67.6\pm1.5$	\\
base+H070p6	&$< 0.230$	&$0.074\pm0.036$	&$67.1^{+1.9}_{-2.1}$	&$< 0.238$	&$0.074^{+0.037}_{-0.036}$	&$67.2^{+1.9}_{-2.0}$
&$< 0.255$	&$0.074^{+0.039}_{-0.037}$	&$67.1^{+1.9}_{-2.1}$\\
base+H072p5	&$< 0.182$	&$0.076^{+0.037}_{-0.036}$	&$67.6^{+1.7}_{-1.8}$	&$< 0.195$	&$0.076\pm0.037$	&$67.6^{+1.7}_{-1.8}$
&$<0.201$	&$0.076^{+0.038}_{-0.037}$	&$67.6^{+1.6}_{-1.8}$	\\

base+H073p02	&$< 0.137$	&$0.078^{+0.035}_{-0.036}$	&$68.2^{+1.4}_{-1.6}$	&$< 0.145$	&$0.079\pm0.037$	&$68.2^{+1.4}_{-1.6}$
	&$< 0.153$	&$0.079^{+0.037}_{-0.036}$	&$68.2\pm1.5$	\\
base+BAO+H070p6	&$< 0.175$	&$0.076\pm0.036$	&$67.7^{+1.4}_{-1.5}$	&$< 0.180$	&$0.075\pm0.036$	&$67.7^{+1.4}_{-1.5}$
&$< 0.187$	&$0.076^{+0.036}_{-0.037}$	&$67.7^{+1.4}_{-1.5}$	\\
base+BAO+H072p5	&$< 0.151$	&$0.077\pm0.036$	&$67.9^{+1.3}_{-1.4}$	&$< 0.160$	&$0.078^{+0.036}_{-0.035}$	&$68.0^{+1.3}_{-1.4}$
&$< 0.168$	&$0.077^{+0.036}_{-0.037}$	&$67.9^{+1.3}_{-1.4}$	\\
base+BAO+H073p02	&$< 0.125$	&$0.079\pm0.036$	&$68.3^{+1.2}_{-1.3}$	&$< 0.135$	&$0.079^{+0.037}_{-0.037}$	&$68.3\pm1.3$
&$< 0.139$	&$0.079\pm0.036$	&$68.3\pm1.3$	\\
\hline
\end{tabular}
\caption{ 95\% CL upper bounds on $\sum m_\nu$ (in eV), mean values and their associated $95\%$~CL errors of
  the reionization optical depth $\tau$ and the Hubble constant parameter $H_0$ (in km s$^{-1}$ Mpc$^{-1}$) for different
  combination of cosmological datasets. The first, second and third
  column show the results for 1, 2 and 3 massive neutrino states,
  respectively. The \textit{base} case refers to the combination of \textit{Planck TT} plus DR9, with
  bias, shot, and a gaussian prior on systematics included.}
\label{tab:tabmnutt}
\end{table*}

\begin{table*}

\begin{tabular}{|c||c|c|c|c|c|c|c|c|c|}

\hline

Dataset &\multicolumn{3}{c|}{\textit{1 massive state}}
&\multicolumn{3}{c|}{\textit{2 massive states}}
&\multicolumn{3}{c|}{\textit{Degenerate spectrum}}\\

\hline

&$\sum m_\nu$ &$\tau$ &$H_0$

&$\sum m_\nu$ &$\tau$ &$H_0$ &$\sum m_\nu$ &$\tau$ &$H_0$\\

\hline\hline

\textit{Planck pol} &$<0.623$ &$0.083^{+0.033}_{-0.034}$

&$65.7^{+3.1}_{-3.8}$ &$< 0.620$ &$0.084^{+0.036}_{-0.034}$

&$65.6^{+3.2}_{-4.3}$ &$< 0.487$ &$0.082_{-0.034}^{+0.035}$ &$65.2^{+2.9}_{-3.8}$ \\

basepol &$< 0.256$&$0.075^{+0.035}_{-0.033}$

&$66.8_{-2.0}^{+1.8}$ &$<0.270$ &$0.075\pm0.034$

&$66.8_{-2.1}^{+1.8}$

&$< 0.276$ &$0.076^{+0.035}_{-0.034}$ &$66.8^{+1.8}_{-2.0}$ \\

basepol+BAO &$<0.176$ &$0.076_{-0.034}^{+0.033}$
&$67.4_{-1.5}^{+1.3}$ &$<0.194$ &$0.076\pm0.033$
&$67.5_{-1.5}^{+1.4}$

&$< 0.185$ &$0.077^{+0.033}_{-0.034}$ &$67.5^{+1.3}_{-1.4}$ \\

basepol+H070p6 &$<0.220$ &$0.077^{+0.033}_{-0.034}$

&$67.0^{+1.7}_{-1.9}$ &$< 0.224$ &$0.075_{-0.033}^{+0.033}$

&$67.1^{+1.6}_{-1.8}$

&$< 0.223$ &$0.076^{+0.033}_{-0.034}$ &$67.1^{+1.6}_{-1.7}$ \\

basepol+H072p5 &$<0.175$ &$0.077^{+0.034}_{-0.036}$

&$67.4\pm1.5$ &$< 0.186$ &$0.075^{+0.035}_{-0.033}$

&$67.5^{+1.5}_{-1.6}$

&$< 0.198$ &$0.076_{-0.034}^{+0.032}$ &$67.1^{+1.6}_{-1.7}$ \\

basepol+H073p02 &$<0.125$ &$0.079_{-0.034}^{+0.033}$

&$67.9\pm1.3$ &$<0.131$ &$0.079^{+0.034}_{-0.033}$

&$67.9_{-1.3}^{+1.4}$

&$< 0.143$ &$0.078_{-0.034}^{+0.33}$ &$67.9\pm1.3$ \\

basepol+BAO+H070p6 &$<0.153$ &$0.076_{-0.034}^{+0.033}$

&$67.6_{-1.2}^{+1.3}$ &$<0.157$ &$0.072\pm0.033$

&$67.6_{-1.2}^{+1.1}$

&$< 0.166$ &$0.077\pm0.033$ &$67.6_{-1.3}^{+1.2}$ \\

basepol+BAO+H072p5 &$<0.135$ &$0.078_{-0.034}^{+0.033}$

&$67.8\pm1.2$ &$<0.140$ &$0.078^{+0.033}_{-0.031}$

&$67.7_{-1.2}^{+1.1}$

&$< 0.149$ &$0.078^{+0.031}_{-0.032}$ &$67.6_{-1.2}^{+1.1}$ \\

basepol+BAO+H073p02 &$<0.123$ &$0.078^{+0.032}_{-0.033}$

&$68.1_{-1.2}^{+1.1}$ &$<0.113$ &$0.079_{-0.034}^{+0.033}$

&$68.0\pm1.1$

&$< 0.124$ &$0.079^{+0.033}_{-0.032}$ &$68.0_{-1.1}^{+1.0}$ \\

\hline

\end{tabular}

\caption{As Tab.~\ref{tab:tabmnutt} but for the \textit{basepol} case, 
  which refers to the combination of \textit{Planck pol} plus DR9,
  with bias, shot, and a gaussian prior on systematics included, see
  text for details.}

\label{tab:tabmnupol}

\end{table*}

\section{Conclusions}\label{sec:conclusion}
The limit found here for the total neutrino mass, $\sum m_\nu<0.183$ at $95\%$~CL, is among the tightest ones in the literature when using the same data sets, and it goes in the same direction than other existing bounds in the literature~\cite{Palanque-Delabrouille:2015pga, DiValentino:2015sam, DiValentino:2015wba, Cuesta:2015iho}. If high-multipole polarization measurements are added in the data analyses, the former $95\%$~CL limit is further tightened ($\sum m_\nu<0.176$~eV). All these findings imply that \textit{(a)} the degenerate neutrino mass spectrum is highly disfavoured by current cosmological measurements; and \textit{(b)} the minimal value of $\sum m_\nu$ allowed in the inverted hierarchical scenario by neutrino oscillation data is discarded at $70\%$~CL. Nevertheless, in the scenario in which the neutrino mass hierarchy turns out to be inverted, a direct measurement of the total neutrino mass from cosmological probes could be fast-approaching.  
If the neutrino mass hierarchy turns out to be normal (as mildly hinted by current results), our neutrino mass limits may tell us something about future directions for searches for neutrinoless double beta decay, $0\nu2\beta$, a rare decay which is currently the only probe able to test the neutrino identity, i.e. the \textit{Dirac} versus the \textit{Majorana} character \cite{Cremonesi:2013vla}. A huge effort has been devoted to assess the sensitivity of future $0\nu2\beta$ experiments~\cite{0n2b,Dell'Oro:2016dbc}, commonly expressed as bounds on the decaying isotope half-life. The latter is related to the so-called effective Majorana mass of the electron neutrino, the $m_{\beta\beta}$ parameter through the relevant nuclear matrix elements (NME). The tightest current bound is $m_{\beta\beta}< 60$~meV, quoted very recently by the KamLAND-Zen
experiment~\cite{KamLAND-Zen:2016pfg}, reaching the bottom limit of the degenerate neutrino mass region. In the normal neutrino mass scheme, future $0\nu2\beta$ experiments would be required to reach a sensitivity in $m_{\beta\beta}$ below $20$~meV, see Ref.~\cite{Gerbino:2015ixa}. Some fraction of next-generation $0\nu2\beta$ experiments could reach that value, being competitive with cosmological bounds and potentially leading to an evidence of $0\nu2\beta$, provided that neutrinos are Majorana particles and that the mixing parameters chosen by nature do not arrange such that $m_{\beta\beta}=0$. In order to achieve these goals and also to perform a succesfull combination of cosmological and laboratory dataset, it is crucial to keep also NME uncertainties under control~\cite{Dell'Oro:2016dbc, Gerbino:2015ixa}.
 Finally, the results reported here show a mild dependence on the neutrino mass scheme choice. Upcoming measurements of galaxy clustering, supported by some robust model comparison, can help enormously in unraveling which scenario describes better the observational findings.

\begin{acknowledgments}
We would like to thank Massimiliano Lattanzi for precious discussions and Luca Pagano for useful comments on the draft.
E.G. is supported by  NSF grant AST1412966.
M.G., S.V. and K.F. acknowledge support by the Vetenskapsr\aa det (Swedish Research
Council). O.M. is supported by PROMETEO II/2014/050, by the Spanish
Grant FPA2014--57816-P of the MINECO, by the MINECO Grant
SEV-2014-0398 and by the European Union’s Horizon 2020
research and innovation programme under the Marie Skłodowska-Curie grant
agreements 690575 and 674896. S.H. aknowledges support by NASA-EUCLID11-0004, NSF AST1517593 and NSF AST1412966.
\end{acknowledgments}

%

\end{document}